\documentclass[final,authoryear,5p,times,twocolumn]{elsarticle}

\usepackage{amssymb}
\usepackage{epsfig}
 \usepackage{subfigure}
\usepackage{graphicx}
\usepackage{rotating}
\def\astrobj#1{#1}

\journal{New Astronomy}

\begin{document}

\begin{frontmatter}



\title{Mt. Suhora M dwarf survey - Detection of eight short-period variable stars}


\author[oan]{L. Fox Machado\corref{cor}}
\ead{lfox@astrosen.unam.mx}

\author[so,isu]{A.S. Baran}

\author[so]{M. Winiarski}

\author[so]{J. Krzesi{\'n}ski}

\author[so]{M.\,Dr{\'o}\.zdz}

\cortext[cor]{Corresponding author. Tel.: +52 6461744580; fax +52
6461744607 }

\address[oan]{Observatorio Astron\'omico Nacional, Instituto de
Astronom\'{\i}a -- Universidad Nacional Aut\'onoma de M\'exico, Ap.
P. 877, Ensenada, BC 22860, M\'exico}

\address[so]{Mt. Suhora Observaroty of the Pedagogical University, ul. Podchor\c{a}\.{z}ych 2, 30-084 Cracow, Poland }

\address[isu]{Iowa State University, 12 Physics Hall, Ames, IA 50011, USA}

\begin{abstract}
The Mt. Suhora M\,dwarf survey  searching for pulsations in low mass main sequence 
stars has acquired CCD photometry of  46 M\,dwarf stars during the first year of the project (Baran et al 2011). 
As a by-product of this search hundreds field stars have been checked for variability. This paper 
presents our initial result of a search for periodic variables in field stars observed in the course of the survey. On 
the basis of the periodicity and the shape of the light curves, eight new variables has been detected, among which
five are $\delta$ Scuti stars and three likely RR Lyrae stars. Although variation in one
of  the stars has been previously detected,  it was classified incorrectly. To support our 
classification, in August 2010, we performed spectroscopic observations to derive spectral types and luminosity classes for 
all eight variable stars.
 
\end{abstract}

\begin{keyword}
stars: $\delta$ Sct, SX Phe -- techniques: photometric, spectroscopic --
stars:oscillations -- stars: individual: \astrobj{TYC\,2644\,832\,1},
\astrobj{TYC\,2594\,1673\,1}, \astrobj{USNO-A2\,1200\,07693323}, \astrobj{TYC\,3527\,2221\,1},
\astrobj{TYC\,3517\,66\,1}, \astrobj{USNO-A2\,1350\,02022405}, \astrobj{BD+32\,3331}, \astrobj{HD\,188524}

\PACS 97.30.Dg \sep 97.10.Ri \sep 97.10.Vm \sep 97.10.Zr \sep
97.10.Sj



\end{keyword}

\end{frontmatter}


\section{Introduction}
\label{sec:int}

One of the advantages of using CCD cameras, compared to photo-electric photometers, is multi-channeling, 
giving opportunity to measure all stars in the CCD's field of view (hereafter FOV). The number of stars may vary from a couple, 
in case of a small FOV and sparse fields, up to thousands for mosaic CCDs pointing near to the plane of the Milky Way. Apparently, 
CCD technique gave us an opportunity to increase our efficiency when an enormous number of stars has to be observed. This is 
particularly important in the sky surveys undertaken recently, like SDSS (York et al 2000), ASAS (Pojma{\'n}ski 2001), and OGLE (Udalski et al. 1994). 
There are also surveys not aiming at covering all the sky, but focusing on specific fields or targets. A good example 
of these is the Kepler mission (Borucki et al. 2010). Some of the surveys are working in the time-series photometry mode, and these are the most suitable 
for variable star observations.

Searching for new variables is also a secondary goal of the survey undertaken by Baran et al. (2011). 
Primarily, they are looking for stellar pulsations in M\,dwarfs with periods of about 40\,min. However, since the 
cadence period is very short (around 10\,sec on average), a flux variation with periods from a couple of minutes up to days, 
in adjacent field stars observed simultaneously, can be easily detected. Here, we report a detection of seven new variable stars and one star with 
a wrong classification found in the literature. Amplitudes of these variability are relatively large so the discovery were mainly done by eyeball. 
Then, the Fourier technique (FT) and the Phase Dispersion Minimization (PDM) were applied to obtain periods of the detected variation.

\section{Photometric data}
\label{sec:obj}

Details on photometry taken in the course of the Mt.\,Suhora survey are given in Baran et al. (2011). 
To date, during the first two years of the survey, they observed  several dozen M\,dwarfs with thousands of field stars. 
All field stars, up to reasonable brightnesses, which were different for each observing night, have been checked for variability. 
Eight stars, \astrobj{TYC\,2644\,832\,1}, \astrobj{TYC\,2594\,1673\,1}, \astrobj{USNO-A2\,1200\,07693323}, \astrobj{TYC\,3527\,2221\,1}, 
\astrobj{TYC\,3517\,66\,1}, \astrobj{USNO-A2\,1350\,02022405}, 
\astrobj{BD+32\,3331} and \astrobj{HD\,188524} show brightness changes on a time scale of a few hours and were identified by eye, although \astrobj{USNO-A2\,1350\,02022405} and 
\astrobj{HD\,188524} show quite small amplitude variations. The list of the new variable stars detected during the survey is presented in Table 1. 
\astrobj{TYC\,3517\,66\,1}, 
was previously known to be a cepheid with a period of 1.585\,d (Akerlof et al. 2000); however, its classification was apparently wrong. Our higher signal 
to noise (S/N) photometry along with spectroscopy revealed that the light curve does not resemble that characteristic for Cepheids. Another star, HD 188524,
was known to be a double star, but no photometry has been published in the literature. We have also searched the International 
Variable Star Index (VSX)\footnote{http://www.aavso.org/vsx/} database  of the American Association of Variable Star Observers (AAVSO) for 
known variables in our fields. None of them, but \astrobj{TYC\,3517\,66\,1}, has been assigned as a variable star.

The light curves of all 8 variable stars are shown in Fig.\,1 through Fig.\,6. In each panel we have indicated the name of the variable, 
the observing band, the date of observations (year-month-day, UT) and the observatory's name. Amplitudes of flux variations range 
from several parts per thousand ({\it ppt}\,=\,0.921\,mmag) for \astrobj{TYC\,2644\,832\,1} and \astrobj{BD+32\,3331}, up to 40\,{\it ppt} for 
\astrobj{TYC\,3517\,66\,1}. 
Minimum to minimum periods, estimated by eye, range from less than an hour for \astrobj{USNO-A2\,1350\,02022405}, up to more than 5\,h for 
\astrobj{TYC\,3517\,66\,1}. 
Periods defined in this way are correct for pulsating stars but it is half of the period for binaries. Small amplitudes suggest that all stars reported 
here might be pulsators, although ellipsoidal variables or small amplitude binaries cannot be ruled out. Time-series spectroscopy could be very handy 
to distinguish between pulsators and binaries.

Not all data have high S/N ratio. Stars presented here were not the target stars in the survey, 
so the exposure times have not been set to obtain high S/N for these stars. If their brightnesses were comparable to those of the targets, we 
could expect good quality data. Otherwise, as in the cases of \astrobj{BD+32\,3331}, \astrobj{USNO-A2\,1350\,02022405} and \astrobj{HD\,188524}, 
a scatter is much higher than 
in other stars having comparable amplitudes of flux variations. More dedicated photometry of those three stars is needed for further analysis.

\section{Period analysis}
\label{sec:obs}

Since the data are sparse, the Fourier technique might not work very well, particularly for data with less than one period cycle. 
Therefore, we have also used Phase Dispersion Minimization (PDM) to better estimate periods.  This tool is very useful for data sets
with gaps or poor time coverage. In some cases PDM gives true as well as multiples of the true  period. 
In those cases we picked the one close to our rough estimation which came directly from the light curve. We calculated periods from 15\,min up to slightly above the expected 
 periods (by rough estimation) .

The results of the PDM application is shown in Fig. 7. The vertical axe of the plots in this figure represents a variance which is a indication of the
goodness-of-the-fit. Any potential periodicities will occur as a minimum variance at a given period.
We derived a 4.34\,hour period for \astrobj{TYC\,3517\,66\,1}. The minimum variance value in this case is close to zero. 
It means that the data are of a very high S/N and there is one 
dominant period present in the data. For two other stars, \astrobj{TYC\,2644\,832}\, and \astrobj{TYC\,2594\,1673\,1}, the minimum variance we derive is also small, 
providing periods 2.80 and 4.35\,h, respectively. \astrobj{TYC\,3527\,222\,1} has two close minima of the variance. It might be an indication of two close periodicities in the data. 
The deepest minimum gives 2.11\,h period. In the case of \astrobj{USNO-A2\,1200\,07693323} there are few local minima of the variance, with the deepest one representing the minimum-to-minimum 
variation. The period associated with this global minimum is 4.19\,h. The Minimum value of the variance for three other stars with the poorest S/N ratios does not look that good as in the 
previous cases. This is caused by big scatter in the data, giving a variance value close to 1.00, even though the period estimation is still correct. That is a natural property 
of the PDM analysis. The periods for \astrobj{BD\,+32 3331}, \astrobj{USNO-A2\,1350\,02022405} and \astrobj{HD\,188524} are 0.88, 0.75 and 1.78\,h, respectively.

\section{Spectroscopic data}
\label{sec:spec}

We conducted spectroscopic observations at the 2.12\,m telescope of the San Pedro M\'artir Observatory in Baja California, Mexico during  August 30, 2010 (UT). 
We employed the same equipment as in Fox Machado et al. (2010).
In particular, we used the Boller \& Chivens spectrograph installed in the Cassegrain focus of the telescope. The Thomson  $2048 \times 2048$ pixel CCD with a 0.14$\mu$m pixel size was 
attached to the spectrograph. The 1200 lines/mm grating was used to cover a wavelength range from 3975 to 5300\,\AA. We employed a dispersion 
of 0.6\,\AA~per pixel (44\,\AA~per mm) with a resolution of 2.6\,\AA. For data reduction we followed the standard procedures using the IRAF package.

Figure\,8 shows normalized spectra of the seven variable stars discussed in the present paper. Because of technical difficulties we were not able to collect a spectrum for
HD\,188524. The derived spectral classifications are indicated in the caption of each panel. We marked characteristic spectral lines of these stars in each panel. 
As can be seen from Fig.\,8, the spectra  of  \astrobj{TYC\,2644\,832\,1},  \astrobj{TYC\,3527\,2221\,1},  \astrobj{TYC\,2594\,1673\,1},  \astrobj{USNO-A2\,1200\,07693323} and 
 \astrobj{TYC\,3517\,66\,1} show characteristics 
of F-type main sequence stars. A-type main sequence stars differ from F-type main sequence stars because in the former the Balmer lines are remarkable for their strength, 
while in the later they are no longer pronounced. In addition, in F-type main sequence stars the weak metallic lines are more numerous and stronger.

A quantitative spectral assignment was made from the intensity ratio of calcium and hydrogen lines 4226\,Ca\,I:4341\,H${\gamma}$, since the 4226\,Ca\,I line gets 
stronger toward later subtypes in F-type stars. We used the ratios of the following lines 4226 Ca I:4481Mg II and 4226 Ca I:4385 Fe I,  which show positive luminosity effect, 
as a luminosity class indicator. Another commonly used feature in the spectral classification 
of the F-type stars is the G-band (lines $\lambda$4305.6, $\lambda$4308) - a molecular band due to the diatomic molecule CH. It appears around type F3, 
depending upon the resolution of the spectrum, and strengthens toward later subtypes. As the figure shows, the G-band is present only in the 
spectrum of  \astrobj{TYC\,2594\,1673\,1},  \astrobj{USNO-A2\,1200\,07693323} and  \astrobj{TYC\,3517\,66\,1} (three panels on the right).

 \astrobj{USNO-A2\,1350\,02022505} shows a spectrum similar to that of  \astrobj{TYC\,2644\,832\,1} --a F0V type star. However, the metallic lines are slightly weaker 
indicating an A9 type, instead. The spectrum of BD+32\,3331, showed at the bottom panel of Fig.\,8, is a typical A0V type star. 
It shows strong Balmer lines accompanied with faint metallic ones.

An additional check of the spectral types was done by comparing the normalized spectra with those of well classified stars available in the literature. 
In particular we used the library of stellar spectra STELIB by Le Borgne et al. (2003), whose spectral dispersion of 1.7\,\AA~per pixel is about twice as low as ours. 
As an illustration, in the Figure\,9 we compared the spectra of  \astrobj{TYC\,2644\,832\,1} (top) and STELIB F0V-type star \astrobj{HD\,90277} (bottom). 
The spectra are quite similar.

\section{Discussion}

Considering the spectral types, periods of the brightness variations and amplitudes of pulsation  
  \astrobj{TYC\,2644\,832\,1}, \astrobj{TYC\,3527\,2221\,1}, \astrobj{BD+32\,3331}
\astrobj{HD\,188524} and \astrobj{USNO-A2\,1350\,02022405} have been classified as
 $\delta$ Scuti-type pulsators .

As well known, 
$\delta$ Scuti-type variables are main-sequence or slightly evolved stars  
located in the lower part of the Cepheid instability strip in the Hertzprung-Russell diagram (H-R diagram). In general, the period range of $\delta$ Scuti stars lies
between 0.5\, to 8\,hours and the spectral types range from A2 to F2.
At present  two main subclasses of $\delta$ Scuti stars are recognized: the small-amplitude subclass and the high-amplitude subclass (Rodr\'{\i}guez \& Breger 2000).
The majority of the known $\delta$ Scuti stars belong to the former subclass. 

The small amplitude $\delta$ Scuti stars (SADS) are population I stars which pulsate with 
a number of nonradial $p$ modes excited to typical photometric amplitude of 0.02 mag, whereas
the high amplitude $\delta$ Scuti stars (HADS)  are mainly Population II monotone radial pulsators with  amplitudes bigger than $\sim$ 0.3 mag. 
While the HADS are low rotators, the SADS  tend to have
much larger $v\,sin\,i$ values, in some cases reaching 200 km\,s$^{-1}$ (Rodr\'{\i}guez \& Breger 2000). 
Since the $\delta$ Scuti-type stars are relatively bright objects, they can be easily detected from the ground. In fact, several $\delta$
Scuti stars have been discovered accidentally when taken as
reference stars of observations of well known $\delta$ Scuti stars
[e.g. Fox Machado et al. 2002; Li et al. 2002, Fox Machado et al. 2007].

In addition to the HADS and SADS there is another  type of pulsating stars which fall in the same region of the H-R diagram as 
the $\delta$ Scuti stars, namely the SX Phoenicis variable stars (SX Phe).
The SX Phe stars are sub-dwarfs with a spectral type A2-F5. Their light 
variations resemble those of the $\delta$ Scuti variables, but with shorter periods from 0.5\,h to 2\,h and larger amplitude of pulsation, up to 0.7 mag in V.
The vast majority of all SX Phe variables are metal-poor stars  and therefore have been found in globular clusters. The field SX Phe stars are rare, even thought are 
interesting targets for the study of radial pulsations (e.g. Fauvaud et al. 2010).

\astrobj{TYC\,2594\,1673\,1}, \astrobj{TYC\,3517\,66\,1} and \astrobj{USNO-A2\,1200\,07693323} are ambiguous cases for classification. 
Regarding the light curve shape these variables  seem to be monoperiodic  radial pulsators, such as Cepheids, RR Lyrae stars, HADS and SX Phe variables. 
In contrast, their amplitudes suggest that all of them might be small amplitude $\delta$ Scuti stars.
Since their period is rather too short for a Cepheid star and too long for a SX Phe variable, these variability types can be ruled out.
Moreover, they cannot be  $\delta$ Scuti stars
by virtue of their late F spectral types.
Therefore, they might either be   
RR Lyrae type stars or small amplitude binaries. In the former case, they could be RR Lyrae type stars of subclass RRc which, as well known,
oscillate in the radial first overtone with a period in a range from about 0.2 to 0.5 days.
In the later case, they could be W Ursae Majoris-type stars (W UMa) which
are close contact binary systems. As it is known,   
the stars in W UMa systems are main-sequence stars  of nearly the same spectral type,  mostly from around middle F to early G. 
The W UMa systems typically have periods between 0.25 and 1.2 days. 

 We tentatively assigned them to the group of 
RRc-type variables, but more dedicated observations are needed for a more convincing classification.

\section{Conclusion}

We have presented our results of a search for new variable stars in field stars observed in the course of the Mt. Suhora M\,dwarf survey project (Baran et al. 2011). 
We have identified a total of eight variable stars. Seven of them are previously uncatalogued variables. To shed more light on the nature of these variable stars, 
spectroscopic observations were performed.

 \astrobj{TYC\,2644\,832\,1}, \astrobj{TYC\,3527\,2221\,1},  \astrobj{USNO-A2\,1350\,02022405}, \astrobj{BD+32\,3331} and
\astrobj{HD\,188524},  have been classified as $\delta$ Scuti-type variables 
taking into account  not only the spectral types but also the oscillation periods and amplitudes of pulsation. 
For these stars photometric observations in a larger time span will probable reveal their multiperiodic behavior.

 \astrobj{USNO-A2\,1200\,07693323}, \astrobj{TYC\,2594\,1673\,1} and \astrobj{TYC\,3517\,66\,1} 
 may be classified as either RRc variables or W UMa binaries. 
 Spectroscopic time-series  could help in disentangling their nature.

\begin{table*}
\caption{A list of new variable stars detected in the course of the Mt.\,Suhora survey. 
Only the star with the asterisk was known earlier but classified wrongly. Coordinates are given for the 2000.0 epoch.}
\begin{tabular}{lcccccr}
\hline
ID & RA  & DEC  & V & Sp. T. & Period  & Class. \\
&(J2000) &(J2000) &(mag)&&(hour)&\\
\hline
\astrobj{TYC\,2644\,832\,1}			& 19:07:47 & 32:35:02 & 10.0 & F0V  & 2.80 & $\delta$ Sct \\
\astrobj{TYC\,2594\,1673\,1}			& 17:04:27 & 32:19:46 & 13.0 & F6IV & 4.35 & RRc\\
\astrobj{USNO-A2\,1200\,07693323} 	& 15:55:18 & 35:18:19 & 12.0 & F6IV & 4.19 &  RRc\\
\astrobj{TYC\,3527\,2221\,1}			& 18:35:40 & 45:40:15 &   9.9 & F0V  & 2.11 & $\delta$ Sct\\
\astrobj{*TYC\,3517\,66\,1}			& 17:36:28 & 49:05:20 & 11.0 & F9V & 4.34 &   RRc\\  
\astrobj{USNO-A2\,1350\,02022405}	& 02:09:14 & 49:27:18 & 13.0 & A9V & 0.75 & $\delta$ Sct\\
\astrobj{BD+32\,3331}				& 19:08:24 & 32:29 13 &   9.6 & A0V & 0.88 & $\delta$ Sct\\
\astrobj{HD\,188524}				& 19:53:42 & 44:36       & 10.4 & F0V & 1.78 & $\delta$ Sct\\
\hline
\end{tabular}
\end{table*}

\begin{figure*}[htb]
\includegraphics[width=14cm,height=15cm]{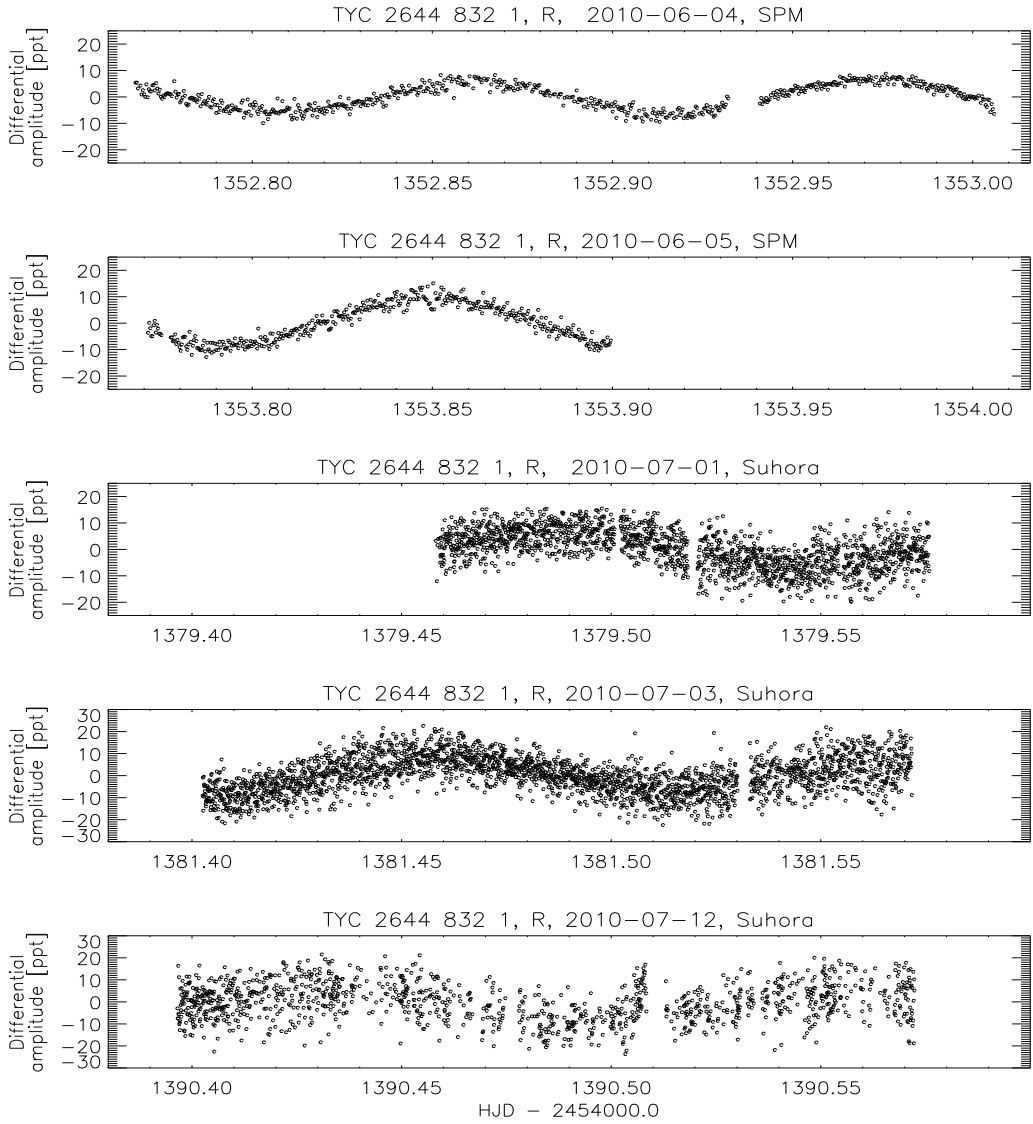}
\caption{Light curves of \astrobj{TYC\,2644\,832\,1}. Data have been taken on five nights. Note different S/N ratios for first two and for the rest of the nights as a 
reflection of the two different telescopes used. The captions in this and the following figures contain the name of the star, the filter used, evening date of observations, 
and site: SPM - San Pedro M\'artir Observatory, Suhora - Mt. Suhora Observatory}
\end{figure*}

\begin{figure*}[htb]
\includegraphics[width=14cm,height=15cm]{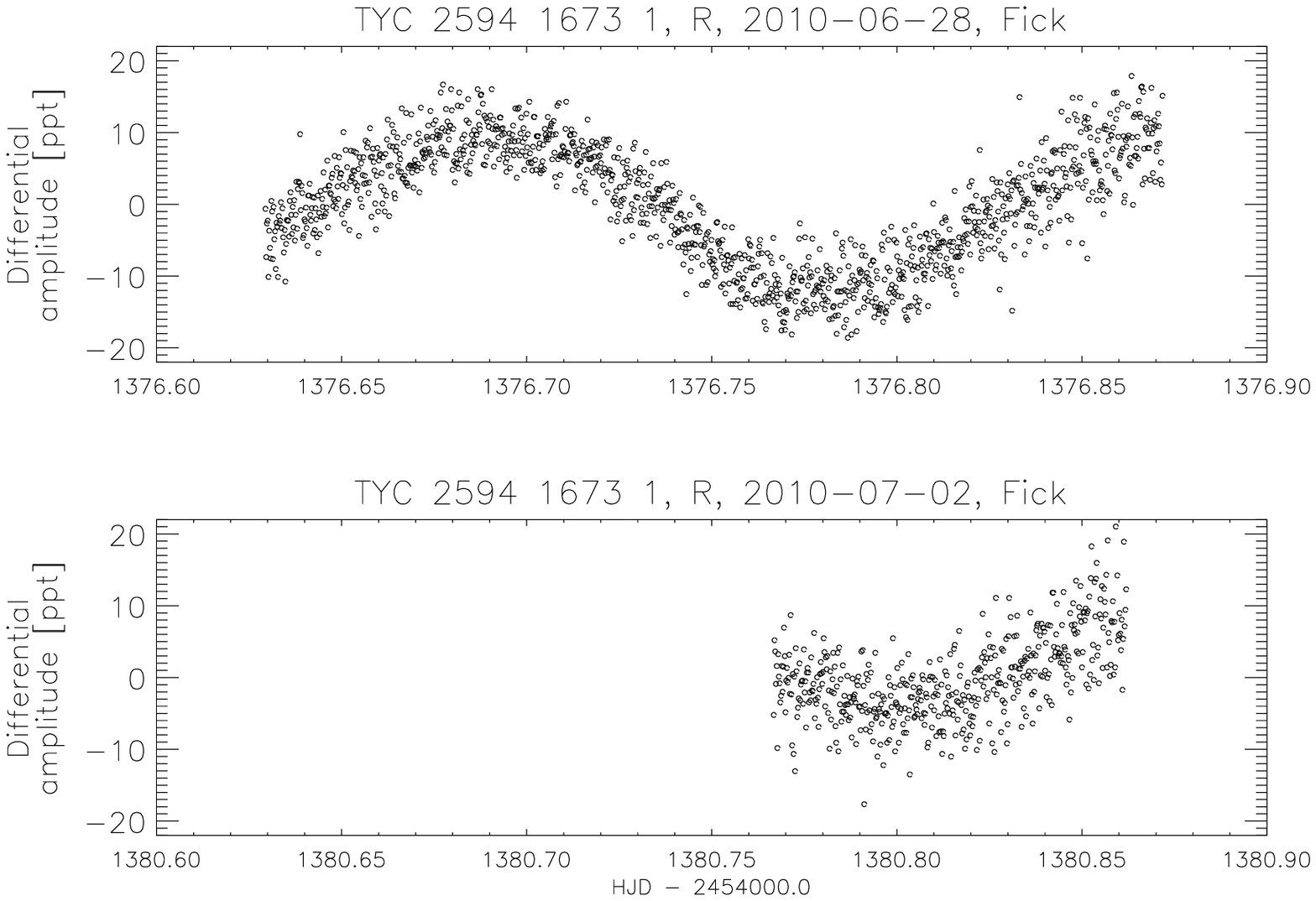}
\caption{Same as in Fig.1 but for \astrobj{TYC\,2594\,1673\,1}.  Fick Observatory. Data have been taken on two nights.}
\end{figure*}

\begin{figure*}[htb]
\includegraphics[width=14cm,height=15cm]{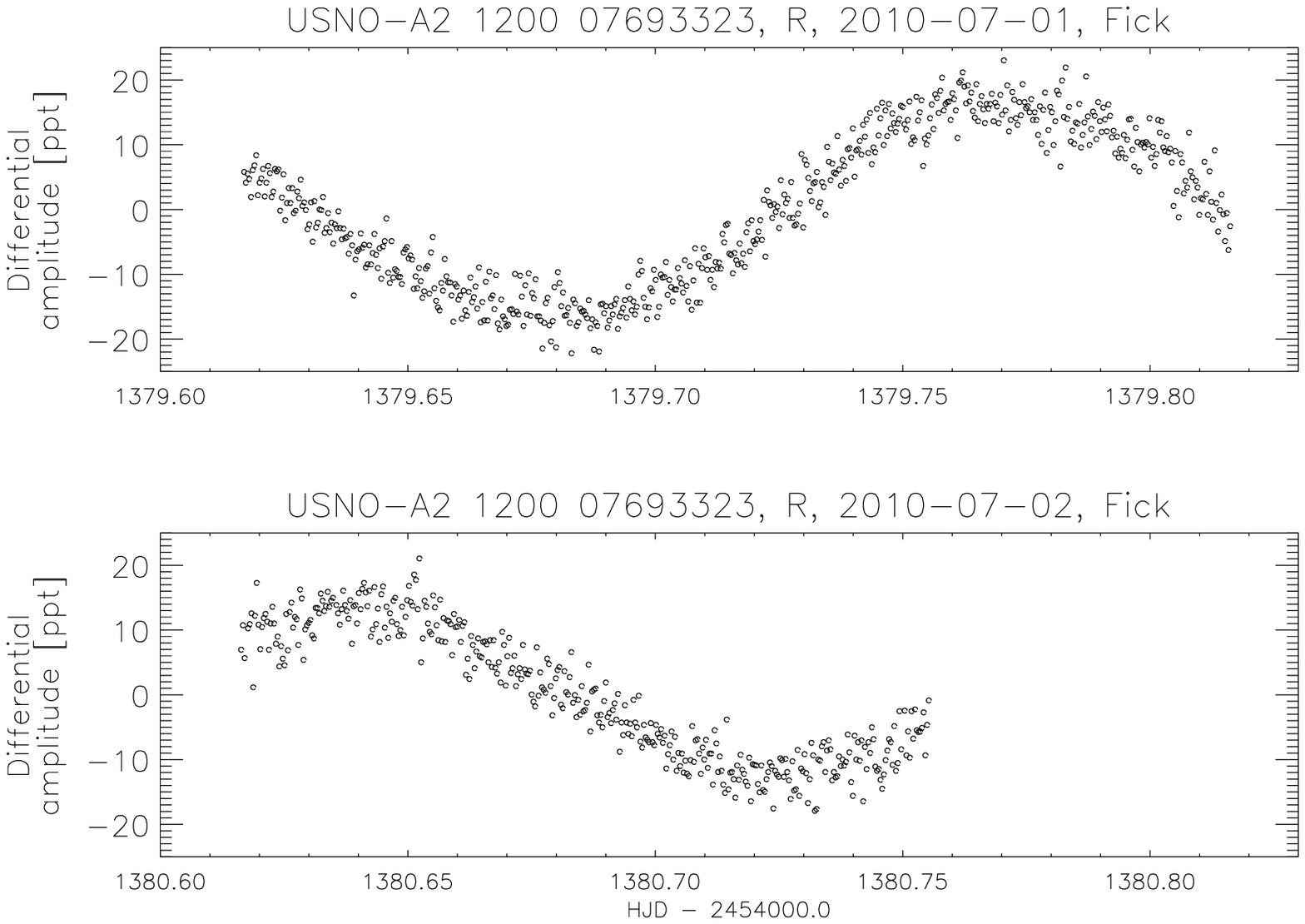}
\caption{Same as in Fig.2 but for \astrobj{USNO-A2\,1200\,07693323}}
\end{figure*}

\begin{figure*}[htb]
\includegraphics[width=14cm,height=15cm]{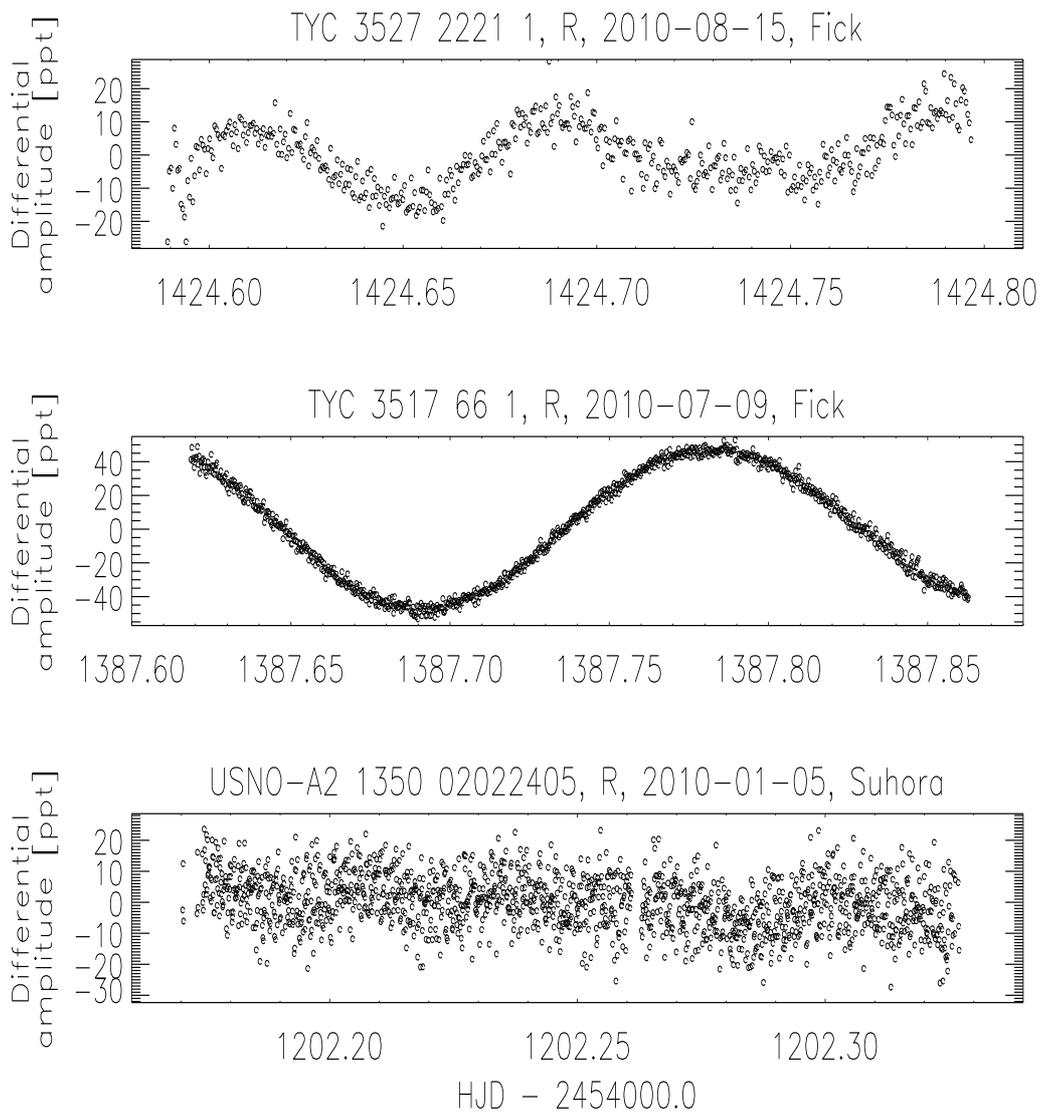}
\caption{Light curves of three variables detected \astrobj{TYC\,3527\,2221\,1}, \astrobj{TYC\,3517\,66\,1},  and 
\astrobj{USNO-A2\,1350\,02022405}.
 \astrobj{TYC\,3517\,66\,1} was previously classified as cepheid.}
\end{figure*}

\begin{figure*}[htb]
\includegraphics[width=14cm,height=15cm]{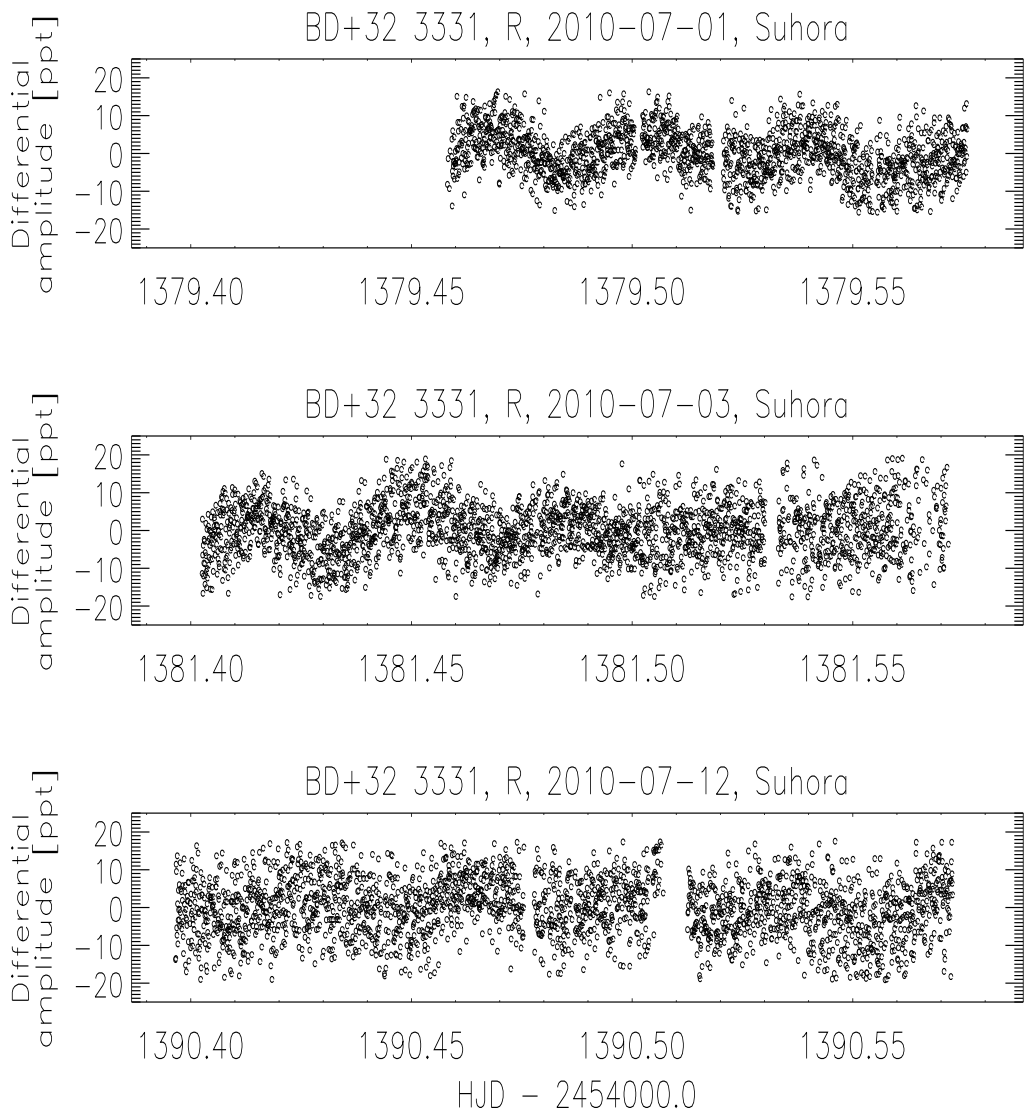}
\caption{Light curves of \astrobj{BD\,+32\,3331}. Data have been taken on three nights.}
\end{figure*}

\begin{figure*}[htb]
\includegraphics[width=14cm,height=15cm]{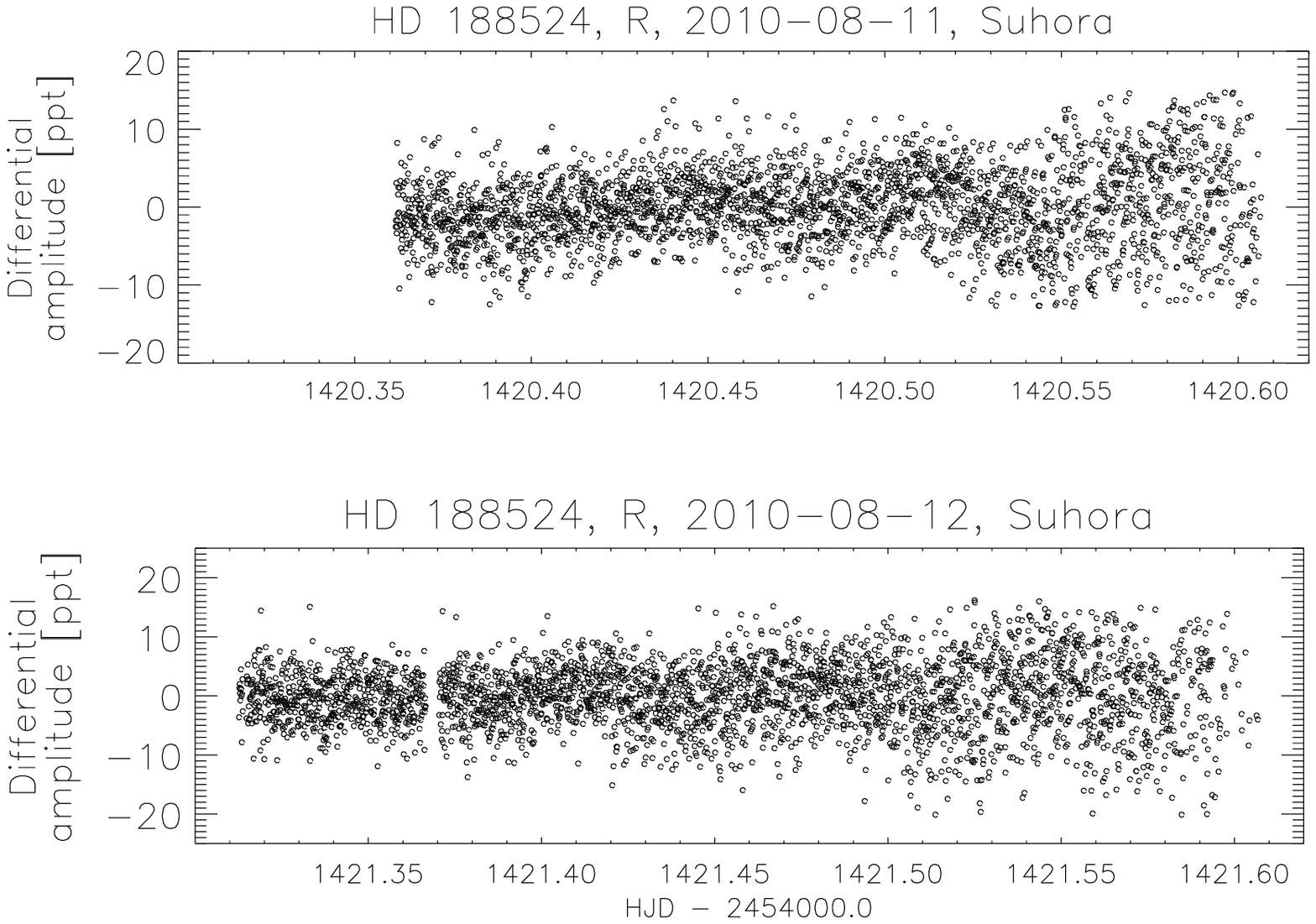}
\caption{Light curves of \astrobj{HD\,188524}.This star has the worst S/N ratio among all stars presented  in this paper because the target star was significantly brighter.}
\end{figure*}

\begin{figure*}[htb]
\includegraphics[width=14cm,height=15cm]{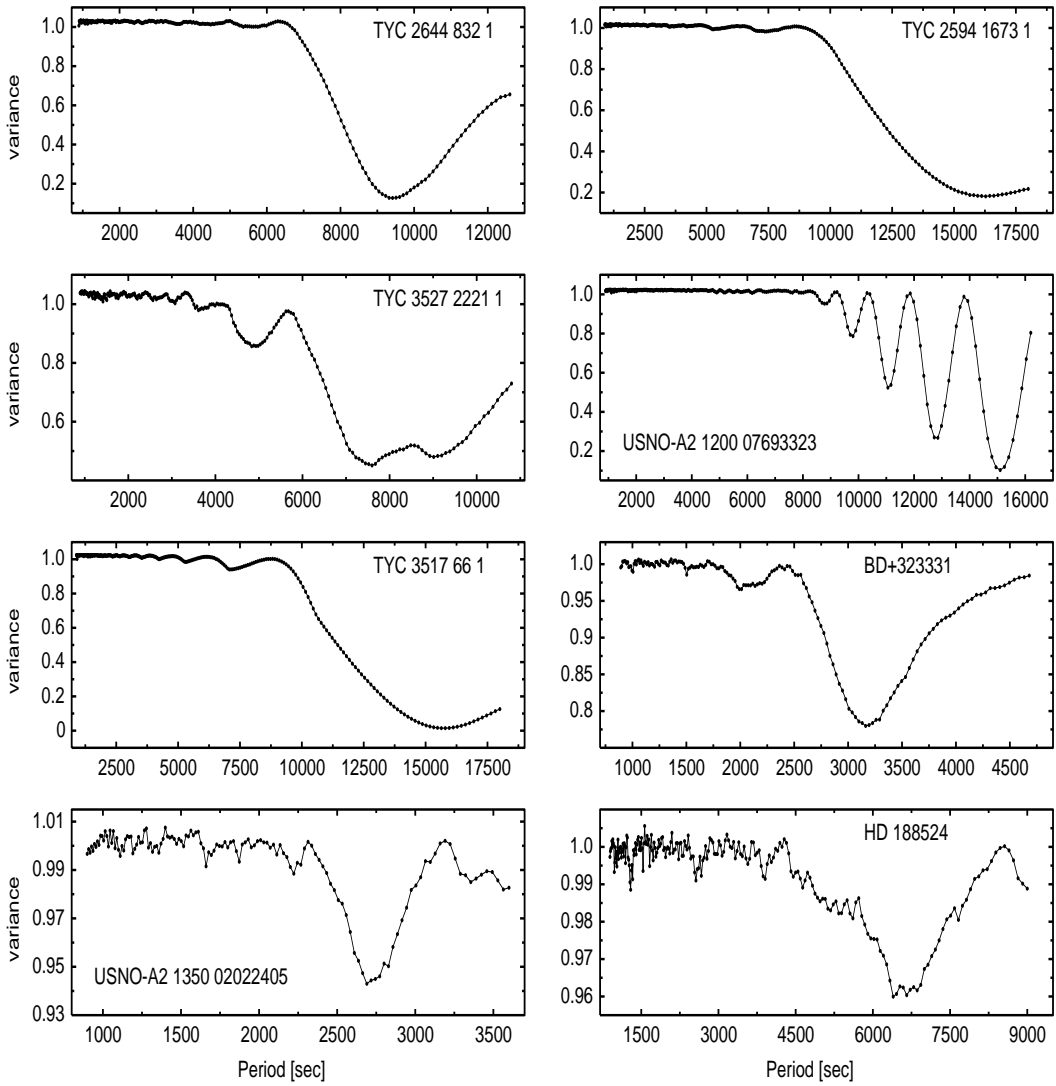}
\caption{PDM for all eight variable stars detected. In some cases we included only one night, even though we collected data on more nights.}
\end{figure*}

\begin{figure*}[htb]
\includegraphics[width=14cm,height=15cm]{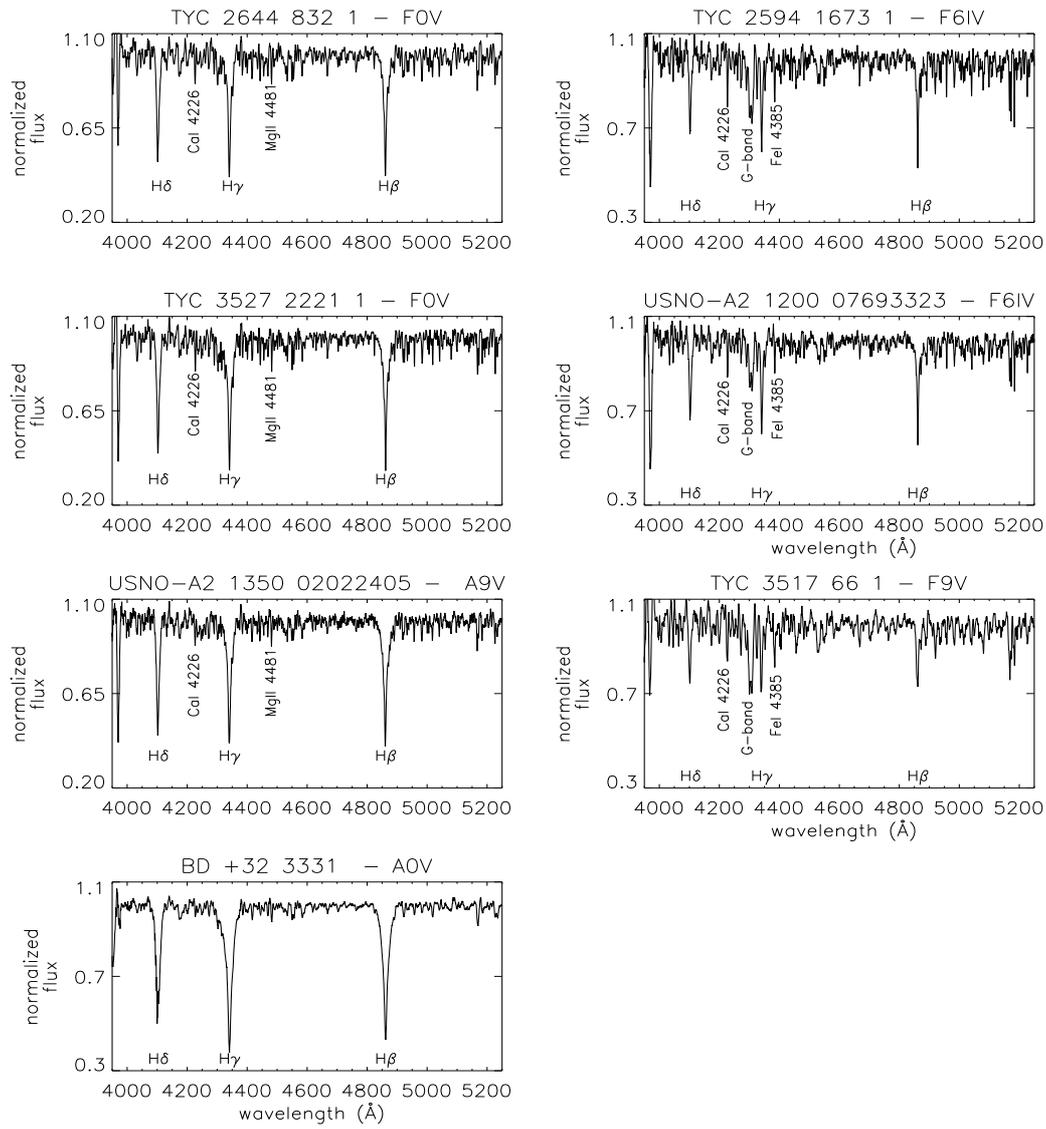}
\caption{Normalized spectra of the seven variable stars. We provide a spectral classification in each panel.}
\label{fig:spectra}
\end{figure*}

\begin{figure*}[htb]
\includegraphics[width=14cm,height=15cm]{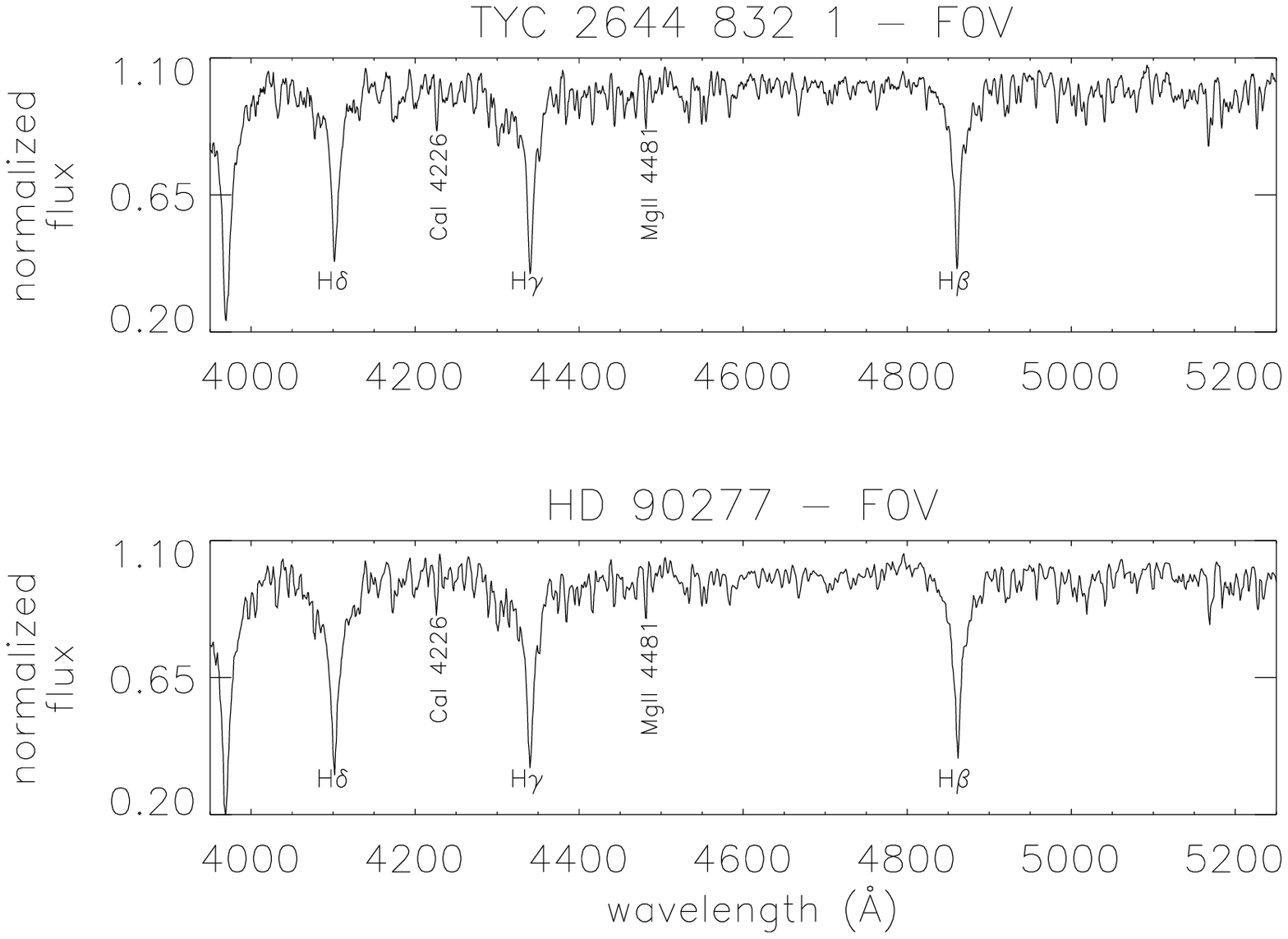}
\caption{A comparison of the \astrobj{TYC\,2644\,832\,1} spectrum with that of a F0V-type star from the STELIB library (see text for details) in the wavelength region between 3950 - 5250\,\AA.}
\label{fig:spectra}
\end{figure*}

\bigskip
{\bf \noindent Acknowledgements}

LFM acknowledges financial support from the UNAM under grant PAPIIT IN114309. This project was supported by Polish Ministry of Science under grant No. N N203 379736. 
We would like to thanks Matthew Thompson from Missouri State University for his support with PDM analysis tool. We would also like to thanks  to the technical staff and 
night  assistants of the San Pedro M\'artir Observatory, Mt. Suhora Observatory and Fick Observatory. We thank W.J. Schuster for a careful reading of this manuscript.

\bibliographystyle{elsarticle-harv}

\end{document}